\documentstyle[psfig,nicV]{article}
\begin{document}
%
%
\heading{%
Reaction rates from Coulomb dissociation: \\
Core excitation effects\\
%
}
\par\medskip\noindent
%
\author{%
Alberto Mengoni$^{1,2}$, Tohru Motobayashi$^{3}$, and Takaharu Otsuka$^{2}$
}
\address{
Applied Physics Division, ENEA, Via Don Fiammelli 2,
       I--40129 Bologna, ITALY
}
\address{
Department of Physics, The University of Tokyo, Hongo, Bunkyo-ku,
       Tokyo 113, JAPAN
}
\address{
Department of Physics, Rikkyo University, Ikebukuro,
       Tokyo 173, JAPAN
}
%
\begin{abstract}
A process, involving the $^{7}$Be core excitation 
in the Coulomb breakup of $^{8}$B into $p + ^{7}$Be 
has been investigated. 
From the experimental results recently obtained in RIKEN we
have derived the mixing amplitude of the 
$| {}^{7}$Be$(1/2^{-}) \otimes \pi(1p_{3/2}); 2^{+}>$ configuration 
in the ground state of $^{8}$B.
Implications on the evaluation of the $S_{17}$ at stellar
energies are discussed.
\end{abstract}
\section{Introduction}

\begin{figure}
\centerline
{\vbox{\psfig{figure=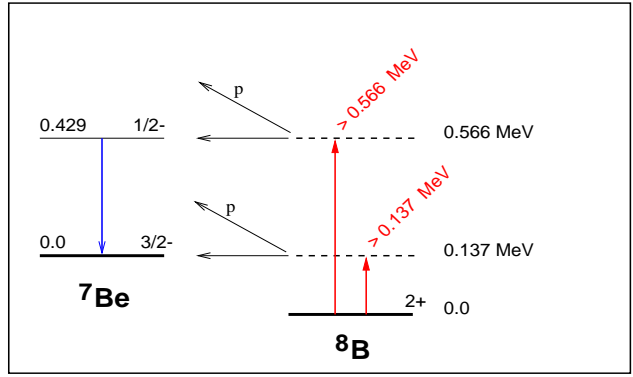,height=5.0cm,width=8.33cm}}}
\caption[]{\small
Breakup channels for $^{8}$B $\rightarrow p + ^{7}$Be 
}
\label{fig1}
\end{figure}

Electromagnetic excitation and/or dissociation
can be induced on unstable nuclei
in experiments employing radioactive ion beams (RIBs).
These experiments allow for a determination of
capture reaction rates involving radioactive nuclei.
There are, however, certain 
cautions which must be taken into account
when applying this method. 
A mechanism which must be considered
is the possibility of core excitation \cite{Me97a}. 
In fact, if a low-lying level of the
core nucleus is present, it can be
excited during the Coulomb break-up process.
As an example, we have analyzed the
reaction
$$
p + {}^{7}\mbox{Be} \rightarrow  {}^{8}\mbox{B} + \gamma 
$$
of crucial importance in solar neutrinos production. 
The Coulomb dissociation of ${}^{8}$B into ${}^{7}$Be$ + p$ 
has been in fact proposed and applied to derive 
the $(p,\gamma)$ reaction rate \cite{Mox94}.
In this case, there is a low-lying $1/2^{-}$ excited state 
in ${}^{7}$Be at 0.429 MeV (see Fig.~\ref{fig1}). 
As well as the ${}^{7}$Be ground state, 
this state can be populated by the $s$-wave
protons in the continuum emitted during 
the breakup process. 
The relative de-excitation
$\gamma$-ray has been recently detected 
and the production cross section derived 
in a breakup experiment in RIKEN \cite{Kix97}.

Here we will present the theoretical interpretation 
of this mechanism together with a quantitative
analysis of the experimental result obtained.
Implications of this analysis on the ${}^{8}$B 
ground-state structure will be discussed and
the relative impact on the calculation of
the $S_{17}$ will be shown.

\section{Coulomb dissociation}
Let us first consider that the Coulomb dissociation
cross section is given by
\begin{equation}
\frac{d\sigma_{CD}}{dE_{x}} = 
\frac{N_{{\rm E}\lambda}(E_x)} {E_x} 
\sigma_{\gamma,p}^{{\rm E}\lambda}(E_x)
\end{equation}
where $N_{{\rm E}\lambda}(E_x)$ is the virtual 
photon number, $E_x$ the excitation energy 
(defined as the sum of the proton-residual nucleus
relative energy plus the proton binding energy)
and $\sigma_{\gamma,p}^{{\rm E}\lambda}(E_x)$
the photo-disintegration cross section. We will
specialize here to dipole radiation, i.e. $\lambda=1$.
The photo-disintegration cross section is given by
\begin{equation}
\sigma_{\gamma,p}^{\rm E1}(E_x) =
\frac{16 \pi}{9} \frac{E_x}{\hbar c} 
\frac{\mu k_{p}}{\hbar^2} \ \bar{e}^2 \
\frac{ 2J_{c} + 1}{2J_{b}+1} \ 
\vert Q^{({\rm E1})}_{b \rightarrow c} \vert^2
\end{equation}
where $\mu$ is the reduced mass of the system,
$k_{p}$ is the wave number of the
proton-residual nucleus relative motion 
in the continuum,
$\bar{e}$ the proton E1 effective charge,
$J_{b}$ is the total angular momentum
of the bound state and $J_{c}$ the spin 
of the residual nucleus in the continuum. 

The photo-disintegration cross section can
be promptly related to the proton capture
cross section (hence to the $S_{17}$) by
detailed balance
\begin{equation}
\sigma_{\gamma,p} = 
\frac{k_{p}^2}{k_{\gamma}^2} \
\frac{ 2J_{c} + 1}{2J_{b}+1} \ 
\sigma_{p,\gamma}.
\end{equation}

The essential ingredients for the calculation of the
Coulomb dissociation cross section 
are the matrix elements
\begin{equation}
{{\it Q^{({\rm E1})}_{b \rightarrow c}}} =
< \Psi _{c} \vert \hat{T}^{\rm E1} \vert \Psi _{b} >
\end{equation}
where $\Psi_{b}$ is the bound-state
wave function and $\Psi_{c}$ the wave function
for the proton in the continuum. The calculation
of these matrix elements is not an easy task, 
in general.
However, they can be readily evaluated under
the particular condition in which the
bound state has a single particle
configuration of type
\mbox{$|{}^{A}\mbox{X}(J_{c}^{\pi})
\otimes \pi(nlj)]; J_{b}>$}.
In this case, they can be decomposed into
the products of three factors
\begin{equation}
Q^{({\rm E1})}_{b \rightarrow c} \equiv
\sqrt{S_b} \ A_{b,c} \ {\cal I}_{b,c} 
\end{equation}
where $S_b$ is the spectroscopic factor of the bound state,
$A_{b,c}$ is a factor containing only angular momentum and
spin coupling coefficients and the radial overlap
\begin{equation}
{\cal I}_{b,c} = \int\limits_{0}^{\infty}
u_{b}(r) r w_{c}(r) dr
\label{eq:roverlap}
\end{equation}
can be evaluated using some potential model for the
calculation of the radial wave functions $u_{b}(r)$
and $w_{c}(r)$. 

The parameters required by these potential 
models are:
(1) the proton-nucleus interaction 
potential in the continuum 
for the calculation of the wave function 
$w_{c}(r)$,
(2) the potential for the calculation of $u_{b}(r)$, and 
(3) the spectroscopic factor $S_{b}$.
It has to be noted, however, that in the
particular case of the 
${}^{7}\mbox{Be}(p,\gamma)^{8}$B reaction,
the proton-nucleus potential in the
continuum does not play any significant
role, as far as the low energy region
(say $E_{p} \leq $ 1 MeV) is considered. 
In fact, it can be easily
shown that even the use of a simple
plane-wave approximation for the
continuum wave function is sufficient
to obtain a good accuracy in the 
calculation of the E1 matrix elements.
Nevertheless, we will use a Coulomb+Woods-Saxon
potential with the parameters fixed by the
bound-state calculation in the evaluation
of the wave function $w_{c}(r)$.

Moreover, one has to consider that
in the calculation of the E1 transition 
matrix elements for loosely bound states
such as the ${}^{8}$B ground state, 
only the {\it asymptotic} behavior 
of the wave function is relevant. 
In fact, a Whittaker function of type
\begin{equation}
u_{b}(r) = \hat{b}_{s} W^{+}_{\eta,l}(k_{b}r)
\end{equation}
has been shown to be sufficient in the
calculation of the E1 matrix elements,
in the low energy region \cite{Xux94}. 
This implies that, in principle, 
the overall normalization
coefficient $\hat{b}_{s}$, 
referred to as {\it asymptotic
normalization coefficient} (ANC), is the
only quantity necessary for the evaluation
of the ${}^{7}\mbox{Be}(p,\gamma)^{8}$B cross section.
The ANCs can be either measured by transfer reactions
or, alternatively, calculated. 
Microscopic and/or potential models
can be used to this end. 
Using our model parameters given below, 
we have obtained $\hat{b}_{s} $ = 0.75 and
0.73 respectively for the $\pi(1p_{3/2})_{1}$
and $\pi(1p_{3/2})_{2}$ single-particle components
(see below for their definition). 
Again we would like to stress here
that the ANCs are more fundamental quantities
as compared to the spectroscopic factors. In fact,
the spectroscopic factor is model-dependent
in the sense that it depends on the
form-factor used in its experimental determination. 
On the other hand, the $\hat{b}_{s}$ only
reflects the {\it asymptotic behavior} of the
wave function and therefore it is independent
on the detail of the interaction which determines
the behavior itself.

\section{Core excitation}
A simple assumption in the ${}^{8}$B ground 
state would consist of the following configuration
\begin{equation}
|{}^{8}\mbox{B};2^{+}> = 
| 3/2^{-} \otimes \pi(1p_{3/2})_{1}; 2^{+}>.
\end{equation}
In our analysis, we will consider an additional
component in the ground state wave
function, namely, we will consider the
following configuration
\begin{equation}
|{}^{8}\mbox{B};2^{+}> = 
\sqrt{1-\alpha^{2}} \ | 3/2^{-} \otimes \pi(1p_{3/2})_{1}; 2^{+}>
\ + \ \alpha \  | 1/2^{-} \otimes \pi(1p_{3/2})_{2}; 2^{+}>
\label{eq:gs2}
\end{equation}
where we have introduced the mixing amplitude $\alpha$. Here,
the two $\pi(1p_{3/2})_{1,2}$ single-particle wave functions 
are identical (derived from the same Woods-Saxon potential),
except that in the second component, the binding energy
is effectively increased by the excitation
energy of the ${}^{7}$Be $1/2^{-}$ state, namely
by 0.429 MeV. The potential parameters we have adopted
are the Barker potential \cite{Ba80}: $r_{0}=1.25$ fm,
$d=0.65$ fm and the well-depth adjusted to
reproduce the proton binding energies
$V_{01}=$-46.54 MeV and 
$V_{01}=$-47.93 MeV respectively for
the two components $\pi(1p_{3/2})_{1,2}$.

A measurement of the Coulomb breakup cross
section for the $^{8}$B into $p + ^{7}$Be 
process has been recently reported in which
the branching ratio for the inelastic channel 
leading to the ${}^{7}$Be low-lying $1/2^{-}$ 
state was measured by detecting the 0.429 MeV
$\gamma$-ray in coincidence with the
breakup events. The result of this experiment
lead to a branching ratio
\begin{equation}
\frac{\sigma_{CD}(1st)}{\sigma_{CD}(gs) + \sigma_{CD}(1st)} = 5\% .
\end{equation}
By calculating the Coulomb dissociation cross
section for both the reaction channels, we
have been able to determine the mixing 
amplitude defined in Eq.~\ref{eq:gs2} as
$\alpha = 0.36$.
This value can be compared with the
theoretical calculation based on a
three-body model reported recently
which gives $\alpha$ = 0.31 \cite{Grx98}.

\begin{figure}
\centerline
{\vbox{\psfig{figure=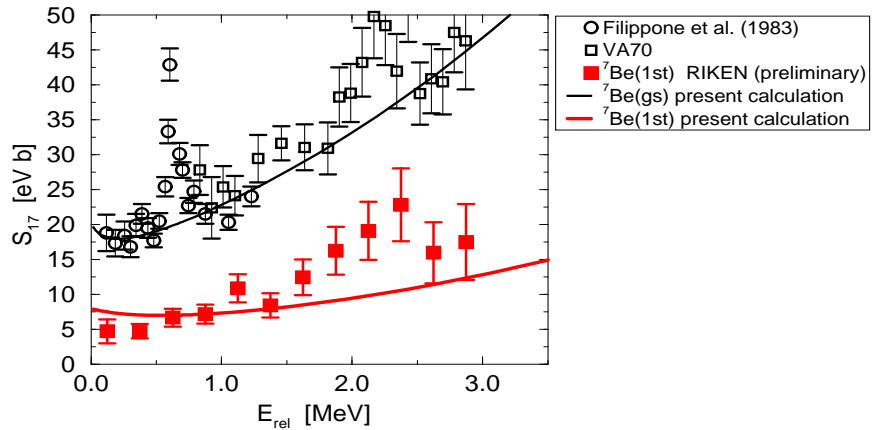,height=6.0cm,width=12.0cm}}}
\caption[]{\small
Astrophysical $S-$factor for the
${}^{7}\mbox{Be}(p,\gamma)^{8}$B reaction. The experimental
values for the inelastic breakup channel are preliminary
results from a RIKEN experiment
}
\label{fig2}
\end{figure}

The corresponding calculation of the
$S_{17}$ are shown in Fig.~\ref{fig2}.
The experimental values for the inelastic
channel are only preliminary.
It can be seen that there is an overall
agreement of the calculation with the
experimental values for both channels.
In turn, this means that our ${}^{8}$B
wave function is reliable. 
In our calculation we obtain
$S_{17}$ = 19.5 eV b
for the elastic and
$S_{17}$ = 9.6 eV b
for the inelastic channel respectively,
at $E_{p}$ = 20 keV.
The value of the inelastic component is
much larger than expected 
and should be seriously considered
in all the breakup experiments aimed at
the determination of this fundamental 
reaction rate.

%
\vfill
\end{document}